\documentclass[12pt]{article}
\usepackage{graphicx}
\usepackage{setspace,amssymb,bm}
\usepackage{cite}
\title{Strength of nanotubes, filaments and nanowires from sonication-induced
scission\footnote{This work was supported by the IRC in Nanotechnology, the EPSRC,
the Gates Foundation and St.~John's College.}}

\author{Y. Y. Huang, T. P. J. Knowles and E. M. Terentjev \\\\
\normalsize{Cavendish Laboratory, University of Cambridge}\\
\normalsize{JJ Thomson Avenue, Cambridge CB3 0HE, U.K.} \\
}
\date{}

\begin{document}

\baselineskip24pt
\maketitle

\begin{abstract}
\noindent We propose a simple model to describe the cavitation-induced breakage
of mesoscale filaments during their sonication in solution. The model predicts a
limiting length below which scission no longer occurs. This characteristic
length is a function of the tensile strength and diameter of the filament, as
well as the solvent viscosity and cavitation parameters. We show that the model
predicts accurately experimental results for materials ranging from carbon
nanotubes to protein fibrils, and discuss the use of sonication-induced breakage
as a probe for the strength of nanostructures.
\end{abstract}

\pagebreak

Measurements of the mechanical properties of nanostructures, and of their
strength in particular, are an essential requirement for fundamental
understanding of the possibilities and performance limits of materials which are
based on such structures. Typically elastic modulus and strength measurements
are performed through mechanical manipulation of individual nanostructures, for
example using scanning probe techniques~\cite{Yu2000b, Wong1997, Kis2002,
Smith2006, Durkan2002, Salvetat1999a, Yang2008a, Guzman2006}; however, because
of the challenges intrinsically associated with nanoscale mechanics, such
measurements remain involved and very time consuming. In this paper, we examine
the fragmentation of filamentous structures under sonication. Based on a
coarse-grained model of this process, we discuss an alternative approach to
probe the strength of elongated nanostructures such as carbon nanotubes, and
show that the limiting length that such structures reach after prolonged
sonication reports accurately on their effective breaking strength. Our results
furthermore shed light on the effect of commonly used sonication treatments on
nanostructured materials.

Sonication is widely implemented in the dispersion of nano- and meso-scale
particles and filaments. The principal origin of the enhanced dispersion is the
ultra-high shear rate attained during cavitation events. Cavitation takes place
when a threshold energy density (estimated by
ref.\cite{Pestman94_CavitationThreshold} to be of the order of
$\sim$10~W/cm$^{2}$) is exceeded by an acoustic compression wave. Recent
experiments and theories highlight the extreme conditions reached during
cavitation: in the vicinity of an imploding bubble shear strain rates up to
$\sim 10^{9}$~s$^{-1}$~\cite{Nguyen97_Soni_TransientFlow,
Hennrich07_CavitationSWNT} and local temperatures of up to
15000~K~\cite{Lohse05_Sonolumin} can be reached. The sonication parameters, such
as container geometry, acoustic power and pulsing rates, determine the frequency
and spatial distribution of bubble creation and implosion events, and therefore
govern the changes of the dispersion morphology with time.

When filaments or tubular particles are being sonicated, a number of studies
have reported unwanted breakage of such particles even in chemically inert
media. Studies performed on carbon nanotubes
(CNTs)~\cite{Hennrich07_CavitationSWNT, Hilding03_CNTDispersion, Chun08_FFF}
report that an initially broad length distribution of CNT lengths changes with
sonication exposure: the mean CNT length gradually becomes shorter, finally
reaching a constant modal length after prolonged sonication. The length
distribution in the final steady-state has been found significantly narrower
than in the initial population of filaments. This observation suggests that the
reduction in filament length, for CNTs at least, was dominated by a
mechanical/shearing process rather than a defect-mediated of random
thermal/chemical breakdown (which would continue to occur without saturation).
In this article we explore a simple theoretical argument for understanding the
observed sonication-induced length reduction and eventual saturation at a given
well-defined short length. We model the shear-induced scission (i.e.
ultrasonication in a chemically inert medium) rather than the shortening
accelerated by chemical effects~\cite{Forrest07_AcidCut}. This model predicts
the existence of a limiting length of a filament, below which no further length
reduction takes place at a given sonication power.

\begin{figure}
\centering \resizebox{0.35\textwidth}{!}{\includegraphics{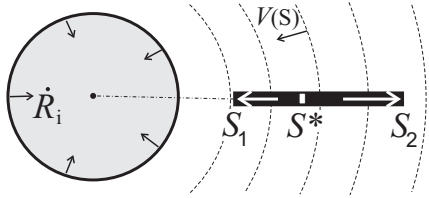}}
\caption{A scheme of cavitation bubble of radius $R_{i}$
collapsing with a wall velocity $\dot{R}_{i}$. The instantaneous velocity field
of the fluid at a distance $S$ from the bubble center falls off with the inverse square
of the distance: $V(S) = R_{i}^2 \dot{R_{i}}/S^2$. The points $S_{1}$ and
$S_{2}$ denote the beginning and end positions of the filament of length $L$.}
\label{Cavitation2}
\end{figure}

We consider a simple potential-flow description of bubble implosion dynamics
which is based on
the radial solvent flow around a bubble; within this framework we then use an
affine estimate to calculate the stress that is exerted on a suspended filament
by the viscous forces transmitted from the solvent. The key parameters
describing this situation are defined in Fig.~\ref{Cavitation2}. We consider a
segment of the filament with a length $L$ and a diameter $d$, which will be
accelerated by the surrounding viscous (surface shear) forces. In a frame moving
with the instantaneous velocity of the filament, the maximum net
tensile stress will act at a point $S^{*}$, which is the fluid stagnation point
on the filament surface. The value of this distance $S^{*}=\sqrt{S_1S_2}$, as
measured from the bubble center, can be found by balancing the
tensile forces $\eta\int_{S_1}^{S^*} V(s)-V(S^*)\, ds = -\eta\int_{S^*}^{S_2}
V(s)-V(S^*)\, ds  $  on both ends of the structure~\cite{Ahir08_CompositeReview}. We
can then evaluate the total tensile force pulling in each direction, and dividing
this force by the cross-sectional area yields the maximal tensile stress
$\sigma_{\rm t}$ exerted on the filament:
\begin{equation}
\sigma_{\rm t} = \frac{8\eta }{d^2} R_{i}^2 \dot{R}_{i}  \left[
\frac{1}{\sqrt{S_{1}}}-\frac{1}{\sqrt{S_{1}+L}} \right]^2.
\label{SoniStress}
\end{equation}
where $\eta$ is the viscosity of the fluid in which the structures are
suspended. For nanostructures with $L\ll S_1$ i.e.\ with a length smaller than
the size of a cavitation bubble, we can expand the square root in
Eq.~(\ref{SoniStress}) to yield: $\sigma_{\rm t}=2 d^{-2}\eta R_i^2 \dot{R}_i S_1^{-3}
L^2$.

The tensile stress $\sigma_{\rm t}$ on the filament decreases dramatically as
the filament length $L$ diminishes, and a characteristic threshold length
$L_{\rm lim}$ below which no scission would occur exists, when the stress is no
longer sufficient to induce fragmentation. The maximal stress occurs for a
filament positioned such that $S_1 = R_i$ (Fig.\ref{Cavitation2}) and therefore we can write the
tensile strength $\sigma^*$ as a function of the limiting length $L_{\rm lim}$:
\begin{equation}
\label{LengthThreshold} L_{\rm lim} = \sqrt{\frac{d^2
\sigma^*}{2\eta (\dot{R}_{i}/R_i)}}.
\end{equation}

Strictly, Eqs.(\ref{SoniStress}) and (\ref{LengthThreshold}) are only applicable
to low-viscosity solvents, as for higher viscosities the probability of
cavitation is diminished and ultrasound energy absorption is increased. For the
case of low viscosity $\eta$, it is convenient to further simplify the equation, by
generically assuming similar $R_i\sim10\;\mu$m, $\dot{R}_i /R_i \sim
10^8~\hbox{s}^{-1}$ values from the literature\cite{Nguyen97_Soni_TransientFlow,
Hennrich07_CavitationSWNT} and $\eta \sim 0.01\,$Pa.s for a typical low-molecular weight
solvent, as quoted above, yielding (in SI units):
\begin{equation}
\label{LengthThresholdReduced} L_{\rm lim} \approx
7\times10^{-4}\, d\,\sqrt{\sigma^*}.
\end{equation}
This is a reasonable approximation for sonication performed in general low
viscosity solvents, provided that cavitation events take place and that the
length of the structures is smaller than the size of the cavitation bubbles. We therefore
suggest that sonication parameters such as pulsing rate, power level (exceeding
cavitation threshold) and container geometry only affect the time at which
$L_{\rm lim}$ is achieved.

We now apply this model to the breakage of filamentous nanostructures such as
multi-walled carbon nanotubes (MWNTs), protein fibrils and silver rods. These
materials are representative examples of, respectively, covalent, non-covalent
and metallic nanostructures. They are assembled through very different
mechanisms: for example CNTs are composed of concentrically rolled graphene
sheets, and protein (amyloid) fibrils consist of an elongated stack of
$\beta$-strands formed as a result of the aggregation of misfolded peptides. We
note that all of our measurements were performed under conditions where the bulk
external temperature of the solvent was kept constant at $\sim 15~^{\circ}$C
using a cooling system, in order to avoid chemical degradation of the structures
from prolonged sonication heating.

\begin{figure}
\centering
\resizebox{0.9\textwidth}{!}{\includegraphics{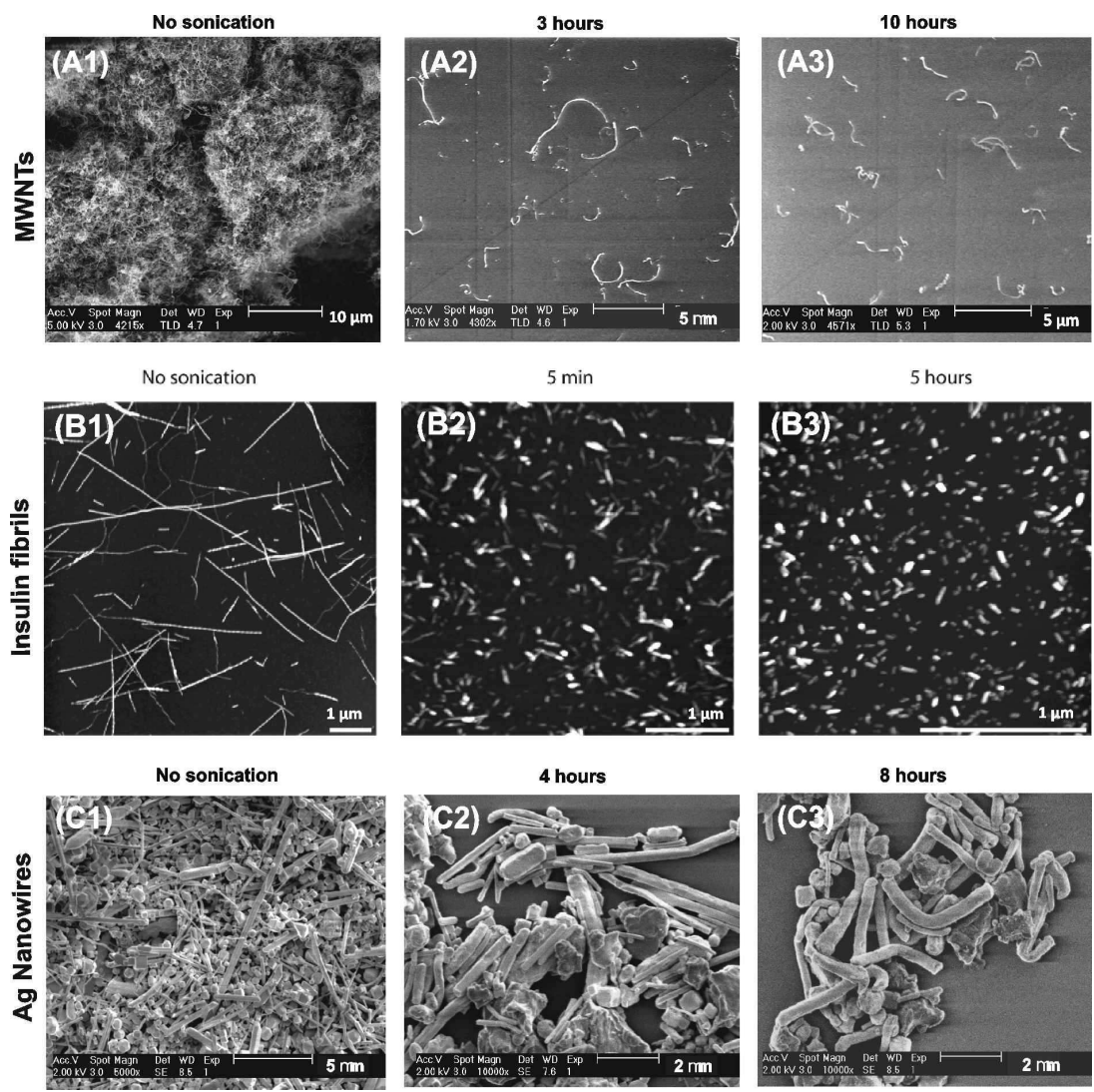}}
\caption{Micrographs of nanostructures before and after sonication. From top to
bottom, A: scanning electron micrographs of MWNTs, B: atomic force micrographs
of insulin amyloid fibrils, and C: silver nanowires.}  \label{SEM_tubes}
\end{figure}

After 3~hr of sonication, the length of MWNTs (diameter 60-100\,nm) was reduced to $\sim$2-6~$\mu$m
as shown by scanning electron microscopy (SEM) in Fig.~\ref{SEM_tubes}(A). This
length range was minimally affected even after an additional 7~hr sonication,
Fig.~\ref{SEM_tubes}(A3), where the few remaining longer tubes were fragmented,
whereas the shorter tubes remain with similar lengths. Assuming $\sigma^* \sim
4$~GPa for CVD MWNTs~\cite{Xie00_CNTStrength}, and considering the widths of the
MWNTs, Eq.(\ref{LengthThresholdReduced}) gives $L_{\rm lim}$ for our
tested MWNTs in the range of 3-5~$\mu$m, in excellent agreement with the
experimental observations in Fig.~\ref{SEM_tubes}(A3). Note that we
are not discussing here the kinetics of the dispersion process, as for instance
the detailed study of~\cite{Huang_prb06}; we are testing the presence and the
value of steady-state $L_{\lim}$.

We next turn our attention to protein fibrils~\cite{Dobson2003, Science2007},
formed here from bovine insulin under conditions which destabilize the native
soluble state of the molecule and promote self-assembly into fibrillar
nanostructures possessing diameters in the range of 3-6~nm and an as-grown
length of several microns. Figure~\ref{SEM_tubes}(B) show the atomic force
micrographs (AFM) for the as-grown, 5~min sonicated, and 5~hr sonicated protein
fibrils. The modal length of the fibrils was reduced by over a factor of 10
within the first 5~min of sonication; however, further sonication for up to
5~hrs only changed the modal length from 130~nm to 70~nm. It is noted that for
insulin fibrils, effective cooling is of key importance for observing the
existence of a limiting length. Sonication performed in the absence of
temperature control frequently leads to complete degradation of structures and
in some cases the formation of amorphous protein assemblies. The tensile
strength of the protein fibrils has previously been estimated from AFM to be in
the range 0.2-1.0~GPa~\cite{Smith2006}. Using the measured values for the
diameter and strength range in Eq.(\ref{LengthThresholdReduced}), yields a value
for $L_{\rm lim}$ between 29 and 130~nm, again in good agreement with the
observations in Fig.~\ref{SEM_tubes}(B3).

\begin{figure}
\centering
\includegraphics[width=\textwidth]{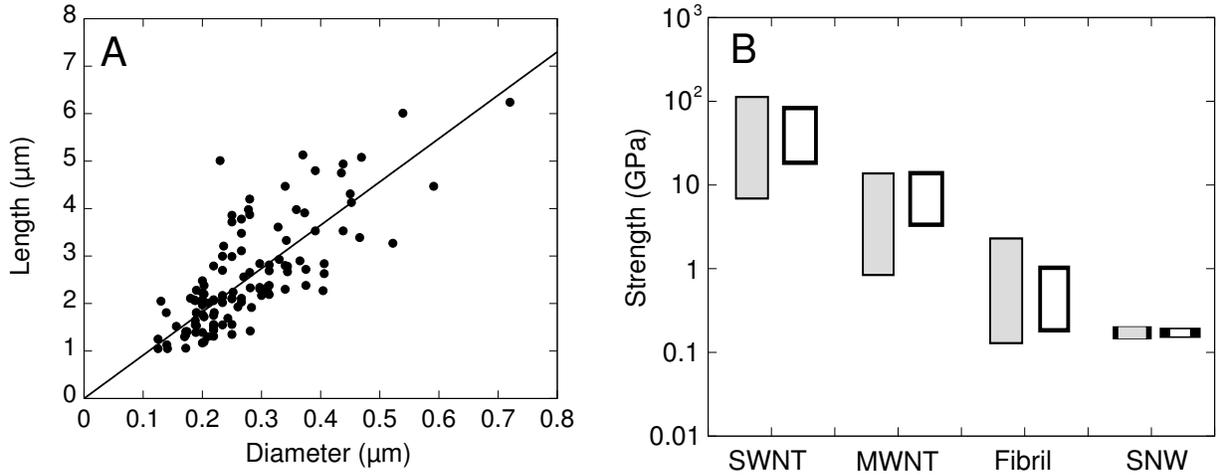}
\caption{A: The terminal length of silver nanowires as a function of their
diameter, fitted to Eq.~(\ref{LengthThresholdReduced}). B:
Comparison between literature values\cite{Wong1997, Smith2006, Salvetat1999,
Li2000, Hennrich07_CavitationSWNT} for the strength of different nano-filaments
(open boxes) and the values obtained in this study using
Eq.~(\ref{LengthThresholdReduced}) (filled boxes). \label{fig:strength}}
\end{figure}

We finally probed the strength of silver nanowires (SNWs) by exposing them to
prolonged sonication. The silver wires, as synthesized (Nanostructured $\&$
Amorphous Materials Inc, Houston USA) had lengths of 10-25 $\mu$m; after 4 hours
of sonication, the length of the structures had been reduced to the range 1-6
$\mu$m and after 8 hours no further reduction in length was observed, Fig.~\ref{SEM_tubes}(C). The
diameters of the silver wires exhibited significant variability, ranging from
100 to 600 nm. This system therefore provides a good basis for probing the
validity of Eq.~(\ref{LengthThreshold}) which predicts a linear relationship
between the diameter of nanoscale filaments and their terminal length under
sonication induced scission. Figure~\ref{fig:strength}A
demonstrates that this linear dependence
is very well satisfied; furthermore, we can extract an accurate estimate of the
tensile strength $\sigma^*$ of the material from the slope of the graph which is
equal to $L_\mathrm{lim}/d = 7\cdot10^{-4}\sqrt{\sigma^*}$ resulting in
$\sigma^*=1.69\pm0.04\cdot 10^8$~Pa. This value is very close to the strength of
bulk silver (170 MPa) \cite{Silver}.

The model was further tested against literature reports for single-wall carbon
nanotubes (SWNTs)~\cite{Hennrich07_CavitationSWNT, Chun08_FFF} and MWNTs from
ref.\cite{Hilding03_CNTDispersion}, and the largest difference between our
estimate for $L_{\rm lim}$ and the measured modal length after scission was
found to be a factor of two for the system of ref.~\cite{Chun08_FFF}. This
finding further supports our previous suggestion that the limiting length is
only weakly dependent on sonication parameters, since all the experiments
referenced above had employed different sonication settings.

Overall, it appears that Eq.(\ref{LengthThresholdReduced}) offers a very
effective approximation, especially considering the simplicity of the model and
the inevitably crude assumptions about the value of tensile strength $\sigma^*$
for some of the cases. We do not exclude the presence of other mechanical
failure mechanisms associated with cavitation. For example, differential
stress-induced bending failure might occur, which is not considered here.
However, tensile failure is promoted since the sizes of the imploding bubbles
($\sim$10~s of microns) are of a similar scale or bigger than the length of typical
filaments. The good agreement between the ultimate length observed in
experiments, and the theoretical $L_{\rm lim}$ calculated based on literature
values of the tensile strengths $\sigma^*$ suggests that tensile failure is the
dominant mechanism of fragmentation. The approach discussed in this paper
therefore can form the basis for a practical evaluation of the tensile strength
of different filaments from the extended sonication-scission experiments,
without the need for micromanipulation of the individual structures. This idea
is illustrated in Fig.~\ref{fig:strength}B, which shows a comparison of the
tensile strength of four different materials computed from their dimensions
after sonication induced fragmentation, and existing values measured in
mechanical experiments. In all cases the experimentally more straightforward
length analysis yields tensile strength values in good /agreement with results
from direct mechanical testing for different types of materials. The accuracy
of our model prediction can be much increased if more precise values of solvent
viscosity and cavitation parameters were to be used for each particular material.

In conclusion we have discussed a simple model which describes the fragmentation
of elongated nanostructures under the action of hydrodynamic stresses imparted
through sonication induced cavitation. We have shown that this model predicts
the existence of a limiting length below which fragmentation no longer occurs.
This length was furthermore shown to be highly dependent on the material
properties of filaments, thereby opening up the possibility of using
sonication-induced fragmentation as a sensitive probe of the strength of a
range of materials in nano-filament form.

\subsection*{Methods}

A Cole Parmer $750$~W ultrasonication system with a titanium tip was used in our
study. The sonication tip pulsed at a 5~s on/3~s off interval, and the output
power level was set at 25\%, yielding an average power density of
$>60$~W/cm$^{2}$ for our container geometry, reliably exceeding the power
density required for cavitation. The MWNTs studied here were obtained from
Nanostructured $\&$ Amorphous Materials, Inc, grown by CVD, with a diameter
range of 60-100~nm and as produced length between 5-15~$\mu$m,
Fig.~\ref{SEM_tubes}(A1). Their dispersion in an organic solvent has been
achieved with the help of pyrene-siloxane (PSi) surfactant synthesized in house.
Amyloid fibrils (Bovine insulin, from Sigma Aldrich) were prepared by incubating
the protein at 60C at the concentration of 10 mg/ml in water adjusted to pH=2 with HCl,
as described elsewhere~\cite{Science2007}. Silver
Nanowires were purchased from Nanostructured $\&$ Amorphous Materials Inc
(Houston USA), quoted grade ($D=$270-330 nm, $L=$10-25\,$\mu$m). Sonication and dispersion were performed in deionized water.


\end{document}